\documentclass[useAMS,usenatbib]{mn2e}

\usepackage{psfig}
\usepackage{graphicx}

\newcommand{\ca}{\ensuremath{$Ca$ $~II$ }}

\title[Chromospheric changes in K stars with activity]{Chromospheric changes in K stars with activity}
\author[M. Vieytes, P. Mauas and R. D\'{\i}az]{Mariela C. Vieytes\thanks{E-mail:mariela@iafe.uba.ar}, Pablo J. D. Mauas, and Rodrigo F. D\'{\i}az
\thanks{Visiting Astronomer, Complejo Astron\'omico El Leoncito operated under 
agreement between the Consejo Nacional de Investigaciones Cient\'{\i}ficas y 
T\'ecnicas de la Rep\'ublica Argentina and the National Universities of La Plata, 
C\'ordoba and San Juan.}\\
Instituto de Astronom\'{\i}a y F\'{\i}sica del Espacio,\\
CC. 67 Suc. 28 (1428)Buenos Aires, Argentina}
\begin{document}

\date{Accepted . Received ; in original form }

\pagerange{\pageref{firstpage}--\pageref{lastpage}} \pubyear{2009}

\maketitle

\label{firstpage}

%\altaffiltext{1}{Member}
%\altaffiltext{2}{Society of Fellows, Harvard University.}

\def\Sc{S$_{\mathrm{Ca II}}$}
\def\teff{\hbox{$T_{\mathrm{eff}}$}}


\begin{abstract}
We study the differences in chromospheric structure induced in K stars by stellar activity, to expand our previous work for G stars, including the Sun as a star.
We selected six stars of spectral type K with 0.82$<B\!-\!V<$0.90, including the widely studied Epsilon Eridani and a variety of magnetic activity levels. We computed chromospheric models for the stars in the sample, in most cases in two different moments of activity.
The models were constructed to obtain the best possible match with the Ca II K and the H$\beta$ 
observed profiles. 
We also computed in detail the net radiative losses for each model to constrain the heating mechanism that can maintain the structure in the atmosphere. 
We find a strong correlation between these losses and \Sc, the index generally used as a proxy for activity, as we found for G stars.
\end{abstract}


\begin{keywords}
 radiative transfer - stars: atmosphere - stars: activity
\end{keywords}



\section{Introduction}
Solar and stellar chromospheric models have been developed to study
the dependency of chromospheric plasma parameters with height and
temperature. \textbf{The best known examples are the models for the
solar atmosphere computed by E. Avrett and his co-workers, in
particular model C for the average quit Sun by \citet{val}, which was later
modified by \citet{fon93}}

In several cases, these models were used to characterize changes due to activity
and spectral type. For example, Kelch et al. (1979) studied a sample
of eight main-sequence stars ranging in spectral type from F0 to M0,
some of which were of similar spectral type and different levels of
chromospheric activity. They computed the photospheric structure
starting from a radiative equilibrium model for the \teff\ of each
star and fitting the \ca\ K line wings. The chromosphere  was built
using the emission core of the \ca\ K line. To  estimate the radiative
cooling rate in the K line they used the K$_{1}$ index \citep{liay78},
which is calculated as the difference between the integrated flux
inside the two K$_{1}$ minima of the \ca\ K line and the corresponding
flux for the model in radiative equilibrium. 

Their results showed that non-radiative heating is important in the lower photosphere of all the late-type stars under study. They found that the value of the K$_{1}$ index and the temperature gradient in the lower chromosphere of these stars, as a function of \teff, divides active and inactive stars, and that the cooling rate in chromospheric lines decreases with \teff. Regarding the chromospheric structure, they found that the temperature minimum moves outward, to lower values of column mass density, with decreasing magnetic activity, {\it i.e.} with decreasing non radiative heating in the lower chromosphere.

Semi-empirical models of the dM star AD Leo in its quiescent state and during a flare were built by \citeauthor{ma94} (1994, 1996). Subsequently, models of two ``basal" ({\it i.e.} inactive) stars of the same spectral type, Gl588 and Gl628, were constructed by \citet{ma97}. 

In a previous paper (\citeauthor{vie05} 2005, hereafter Paper I), we
computed chromospheric models for a sample of dwarf stars of spectral
type G, including the Sun as a star, \textbf{using the FAL
models \citet{fon93} as a starting point}. Our purpose was to study the
changes in chromospheric structure induced by magnetic
activity. The stars we modeled were chosen to have similar colors than
the Sun, and therefore similar photospheric structures, but different
chromospheric activity levels, probably due to different ages and/or
rotation periods. These stars can be considered as solar analogues,
since they share several characteristics with the Sun.  

To extend our research to cooler stars and to study how the chromospheric structure changes with spectral type and chromospheric activity, in this paper we perform a study similar to the one in Paper I for several dwarfs of spectral type K, selected with similar colour, {\it i.e.} similar photospheric structure, and with different levels of magnetic activity. 

As the base for our sample we selected one of the most studied K stars, Epsilon Eridani (HD 22049), which is an active star of 
spectral type K2 V ($B\!-\!V$=0.88), with $T_{\mathrm{eff}}=5110$~K. 
This star has been widely studied because it is one of the ten nearest stars. It has two planets 
and a belt of dust particles around it, which has been compared
to the Kuiper belt in the Sun. These facts make this stellar system resemble our own Solar System. 

Several chromospheric models have been computed for this star. 
\citet{kel78} modeled the lower chromosphere to match the Ca II K line profile and integrated fluxes of the Mg II h and k lines. 
Using observations of the ultraviolet lines of C II, Mg II, Si II and Si III from the IUE satellite, 
\citet{simo80} obtained a model for Epsilon Eridani, which also reproduces hydrogen line profiles not fitted by Kelch's model. The thermal structure of this model has the onset of the transition zone deeper in the chromosphere and a lower temperature in the plateau than Kelch's model.

Another chromospheric model for Epsilon Eridani is the one by \citet{that91}, who 
fitted the Ca II K line, the infrared triplet lines of Ca II, the Na D doublet, 
H$\alpha$ and H$\beta$.
More recently, \citet{sim05}, using ultraviolet 
observations from STIS and FUSE, developed a new semiempirical model for 
the upper chromosphere and lower transition region of this star keeping the photosphere and lower 
chromosphere of \citet{that91}. 

Finally, \citet{nes08} studied the relative element abundances from the conora and upper transition region of Epsilon Eridani, using observations from Chandra, EUE, FUSE and XMM-Newton.

This Paper is arranged as follows: we present our stellar sample and discuss the observational 
data in \S 2. In \S 3 we describe the models and show the results. In \S 4 we compute
the energy requirements to sustain the chromosphere, and compare the results with those obtained for G stars in Paper I. Finally, in \S 5 we discuss the results.

\section{Our stellar sample}

%The stars in our sample were selected with $B\!-\!V$ colour index close to Epsilon Eridani, in the range 0.82$<B\!-\!V<$0.90, and with different levels of magnetic activity. 

\begin{table*}
 \centering
 \begin{minipage}{160mm}
  \caption{The stellar sample. Columns 3 to 6 list the stellar parameters (from Perryman et~al. 1997, and from Cincunegui \& Mauas 2004). The next three 
  columns give the \Sc\ measured by Henry et al. (1996) at CTIO and by
  Cincunegui \& Mauas (2004) at CASLEO, both converted to Mount Wilson \Sc\ \textbf{compared whit \Sc\ calculated from our models}; and the last two columns
  list the observing dates for each spectrum we used.} 

 \begin{tabular}{@{}lllllllllll@{}}
  \hline
  HD ~(Name) &  \emph{S. type} &  $B\!-\!V$ &  \teff (K) &  $[$Fe/H$]$ &  $S_{\mathrm{CTIO}}$ 
 & $S_{\mathrm{CM}}^{max}$/\textbf{$S_{\mathrm{mod}}^{max}$} &  $S_{\mathrm{CM}}^{min}$/\textbf{$S_{\mathrm{mod}}^{min}$}  & Min & Max \\ \hline
 17925 (V* EP Eri)      & K1 V  &  0.86 &  4956 & 0.10 & 0.662  &  0.792/\textbf{0.584} & 0.566/\textbf{0.520} & 11/22/02 & 12/5/03   \\ 
 22049 ($\varepsilon$ Eri)  & K2 V  & 0.88 & 5110 & -0.14 & 0.483  & 0.555/\textbf{0.468} & 0.440/\textbf{0.389} & 11/21/02 & 3/9/04    \\ 
 26965 (V* DY Eri) & K1 V  &   0.82 & 5203 &  -0.25    & 0.185  &  0.188/\textbf{0.149} &   0.138/\textbf{0.147}  &  3/9/04  &  8/11/00  \\ 
 37572 (V* UY Pic)  & K0 V  & 0.85 & 5175 & ---   & 0.952  &  0.703/\textbf{0.687} &   0.691/\textbf{---}   &  --- & 11/24/04 \\ 
 128621 ($\alpha$ Cen B) & K1 V  &   0.90 & 5037 & 0.24 & 0.209 &  0.247/\textbf{0.180} & 0.164/\textbf{0.139}   &  8/13/00 & 9/11/03  \\ 
 177996 (---) & K1 V  &  0.86 &  5092 & ---  & 0.861  &  0.821/\textbf{0.798} &   0.613/\textbf{---}  &  --- &  6/27/02  \\ 

\end{tabular}
\end{minipage}
 \label{tab:ch}
\end{table*}

The largest observational study of chromospheric activity is the one started in 1966 at the Mount Wilson Observatory, which at present includes more than 2200 stars in the spectral range between F and early K. 
As the indicator of chromospheric stellar activity, they use the \Sc\ index, which is the ratio of
the fluxes in the H and K line cores and two nearby reference windows 20~\AA~
wide \citep{vau78}. The emission in the cores of these lines increase with increasing 
chromospheric activity, $\it{{\it i.e.}}$ with increasing surface magnetism. In this work we used the same activity indicator.

To select the stars in our sample, we require that $0.82<B\!-\!V<0.90$, a colour similar to $\varepsilon$ Eri, and that the magnetic activity levels are different. All the stars are part of the library of southern late-type dwarfs published by \citeauthor{cin04} (\citeyear{cin04}, hereafter CM04). 

The stellar parameters of the stars in our sample are listed in Table~\ref{tab:ch}. 
In the third column we list the 
spectral type, in the fourth to sixth columns we indicate the colour index
$B\!-\!V$, \teff\ and the metallicity. 
In column 7 we show the mean values of the \Sc\ index obtained at the Cerro Tololo InterAmerican Observatory \citep{he96}, 
and in columns 8 and 9 the maximum and minimum \Sc\ obtained from our spectra (see Cincunegui et al. 2007 for details on how this index is obtained) \textbf{and from the models we built in this paper}.
Finally, in the last two columns of Table~\ref{tab:ch} we 
include the observing dates of each spectrum used in the present work.

The observations were made at the 2.15 m telescope of the Complejo Astronomico El Leoncito (CASLEO), located in San
Juan, Argentina. They were obtained with a REOSC spectrograph designed to work 
between 3500 and 7500 \AA\, and a 1024 x 1024 pixel TEK CCD as detector.
The spectral resolution ranges from 0.141 to 0.249 \AA\ per pixel
($R=\lambda/\delta\lambda \simeq 26400$). We refer the reader to CM04 for more details on the observations and the data reduction. 

%We computed an \Sc\ index from our spectra according to the following procedure. 
%In CM04 we included in our sample 18 of the
%``calibration stars'' listed by \cite{he96}~(1996), which are stars with more than 100 observations 
%in the Mount Wilson survey and with activity levels almost constant. Typically we observed each one 
%of these stars between 8 and 12 times, and calibrated the spectra in flux. Then we measured an
%$S'$ index defined as the ratio between the weighted fluxes in two passbands centered in the H 
%and K lines, using a weighting 
%function that mimics the instrumental profile of the Mount Wilson instrument, and the average 
%flux in two continuum passbands nearby. For each calibration star, we averaged 
%its $S'$ values, and we fit these mean values with the Mount Wilson ones. Finally, we converted the $S'$ indexes to \Sc\ using this calibration.

For all the stars, we have several spectra obtained in different observing runs. 
To study the differences in atmospheric structure with activity level, in this paper we consider, in most cases, two spectra for each star, chosen 
between those with the better signal to noise ratio.
Generally we selected the spectra showing the lowest and the highest levels of activity, except for 
HD 177996 and HD 37572, for which the least active spectra are very similar to the most active ones of HD 22049 and HD 17925 respectively. In this way, we built 10 different models. It is important to note, given the dependence of activity level with the observation time, that all the line profiles used to build the models are simultaneous.

\begin{figure}
\centering
\includegraphics[%
  clip,
  scale=0.4]{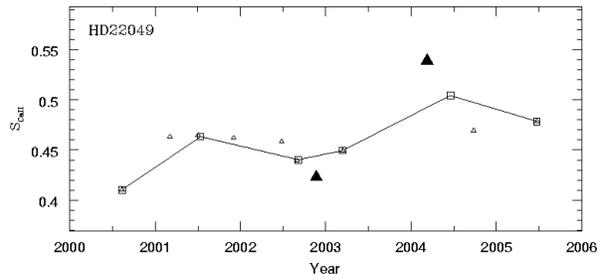}
\caption{\rm \Sc\ for each observation for $\varepsilon$ Eri in our library. The open triangles are the values for the different observations, the squares indicate the annual averages, and the largest full triangles show the two spectra modeled in this paper.}
\label{ciclo}
\end{figure}

In Figure~\ref{ciclo} we show the \Sc\ index of $\varepsilon$ Eri
obtained from our observations (open triangles). The two spectra
modeled in this paper are  
represented by full triangles. The difference in the Ca II K line flux between the maximum and
minimum is 17\%. With squares we also present the annual average of
the \Sc\ index. For details on the variability of $\varepsilon$ Eri,
see \citet{bu}. 

%______________________________________________________________

\section{The chromospheric models}
For each star we built a different chromospheric model, assuming one-dimensional, plane-parallel atmospheres. 
We simultaneously solved the equations of hydrostatic equilibrium, radiative transfer
and statistical equilibrium, using the computer code Pandora. A description of
this code can be found in \citet{av03}. 

For a given  distribution of temperature with height, we
self-consistently computed non-LTE populations for 15 levels of H, 13
of He I, 6 of He II, 15 of Fe I, 8 of Ca I, 5 of Ca II, 7 of Mg I, 6
of Mg II, 21 of Si I, 8 of Na I and 6 of Al I. The atomic models we
used for H and Ca II are described in \citet{ma97} and
\citet{fa98}. The \ca\ lines and Ly$\alpha$ were computed using
Partial Redistribution, \textbf{as it has been done in previous
chromospheric models (like, for example, the \citet{val} solar models) }.

An important element to include in this kind of modeling, in particular for the coolest stars, is the effect of bound-bound absorptions due to the numerous atomic and molecular lines present in the stellar atmosphere, referred to as line blanketing \citep{fa98}, which plays a crucial role in determining both the emergent energy distribution and the physical structure of the atmosphere. In solar-type stars the most important effects come from neutral or single ionized metals. In even cooler stars, molecular bands, as CN, CO, H$_{2}$O, etc, could dominate. 
In this paper, line blanketing is treated in non LTE, as explained in
\citet{fa98}, assuming  
the source function is given by

\begin{equation}
S_{\nu}=\alpha J_{\nu} + (1-\alpha)B_{\nu}, 
\end{equation}
where $B_{\nu}$ is the Planck function and $J_{\nu}$ is the mean intensity. $\alpha$ is the 
scattering albedo, for which we used the expression given by \citet{and89} which depends on wavelength, depth and temperature. 

From the finished model, we 
computed the emitted profiles of H$\beta$ and of the Ca II H and K lines, and modified
the model until we found a satisfactory match with the observed profiles. As a check of the accuracy of the
models, we also compared the observed computed profiles of the Mg I b and the Na I D lines for each 
model (details of these features can be found in \citeauthor{ma88} \citeyear{ma88} and \citeauthor{di07} \citeyear{di07}).

For comparison with synthetic profiles, the observations were converted to the stellar surface 
through

\begin{equation}
log(F_{surf}/f_{earth})=0.35+0.4 (V+BC)+4 log(T_{eff}), 
\label{flux}\end{equation}
where $F_{surf}$ is the stellar surface flux, $f_{earth}$ is the flux observed at earth,
V is the visual magnitude, BC is the bolometric correction given by \citet{joh66}, and 
$T_{\mathrm{eff}}$ is the effective temperature for each star, given in Table \ref{tab:ch}.

Of course, semiempirical models like this one are only a first
approximation to the structure of stellar chromospheres, which are
neither static nor homogeneous. Regarding temporal variations, we took
care of picking our observations at times when no flares were
present, using the method explained in \citet{CDM07}. 
Spatial inhomogeneities characteristic of magnetically active stars,
like starspots or active regions, cannot be resolved on the stellar
surface. The models presented here, however, can be used as a first
step to build two component models as was done, for example, by \citet{mafa96}. 

Faster temporal variations, like waves, cannot be reproduced with this
kind of models, of course. We are also not considering possible
small-scale spatial inhomogeneities like, for example, the
chromospheric bifurcation proposed for the Sun by \citet{Ayres81}, 
which should be produced by CO cooling. However, on one hand this
cooling was probably overestimated \citep{MAL90}, and on the other
it is probably too slow compared to atmospheric dynamics
\citep{WBS07}. In any case, homogeneous models provide information on
the "mean" state of the stellar atmosphere, where the different
components are weighted by their effect on the emitted radiation, in
particular on the spectral features under study.

\subsection{Stellar parameters for $\varepsilon$ Eri}
Before building the model atmosphere, a set of atmospheric parameters has to be determined.
Both the surface gravity and the metallicity are fundamental input parameters in any atmospheric 
model, and the effective temperature, although is not needed as input, is used in Ec. \ref{flux} to calculate the stellar surface flux needed to analyze the results.

In Table~\ref{tab:g} we summarize several values of these quantities that can be found in the literature. Given the astrophysical interest on $\varepsilon$ Eri, \citet{dra93} recognized the necessity of determining these parameters with high precision and they summarized the methods used to obtain them until 1993, and the validity of these determinations. To improve these values, they determined the surface gravity, metallicity and effective temperature in a self-consistent way, comparing the equivalent widths of several Fe I, Fe II and Ca I lines with theoretical profiles from different model atmospheres. 
The parameters derived by \citet{dra93} were used in the most recent model for $\varepsilon$ Eri by \citeauthor{sim05} (\citeyear{sim05}, hereafter SJ05), although they recognized that the value of \rm{log(g)} adopted could be too high (private communication).

The difficulty in the calculation of the surface gravity is that it is indirectly determined from the values of mass and stellar radius. Since these two parameters can be calculated more precisely for stars in binary systems, we studied another star of our sample, $\alpha$ Centauri B (HD 128621), pertaining to the system $\alpha$ Centauri AB. For close systems like this visual binary, the stellar radii and masses can be derived with an error of 1 to 10\% \citep{guedem}.

%In order to decide which of these values of gravity are a better selection, we have studied another star in our sample, $\alpha$ Cen B (HD 128621). As this star belongs to a multiple system,it is possible that its surface gravity may be calculated with more exactitude than $\varepsilon$ Eri.
According to \citet{cay01}, the values of \rm{log(g)} found for $\alpha$ Cen B range from 4.51 to 4.73, with an average value of 4.60. We therefore adopted a value of $log(g)=4.65$ for all the stars in our sample, since this value is contained in the range given by \citet{dra93}, considering the error in their calculation ($log(g)=4.75\pm0.1$). \textbf{This same value of \rm{log(g)} was adopted by \citet{nes08} in their recent study of the corona and transition region of $\varepsilon$ Eri.}

Regarding the rest of the stellar parameters, we adopted \teff\ = 5110 K \citep{tom99}, which is close to the value by \citet{dra93}. We adopted solar metallicity as a good approximation for $\varepsilon$ Eri, as has been done in all the previous models for this star, since it is a young star which is probably not metal defficient. This was suggested by \citet{kri66}, who built a grid of model atmospheres for $\varepsilon$ Eri with different metallicities to fit the \ca\ K line and found that using solar metallicity results in the best agreement with observations. 

In the case of $\alpha$ Cen B, \citet{ay76} built two models for this star assuming in one case solar metallicity and in the other an abundance twice as large. They concluded that the computed profiles of the \ca\ K line differ very little and are in both cases consistent with the observations.

%\textbf{Nevertheless, we changed the metallicity in our model to the value calculated by \citet{zha02} for $\varepsilon$ Eri, following the value adopted by \citet{nes08}. We have not found any significant difference than when we use solar metallicity}

\begin{table} 
 \centering
 \begin{minipage}{86mm}
 \caption{Stellar characteristics for Epsilon Eridani (HD 22049)
from \citet{cay01} and Table 1 from \citet{dra93}.}

  \begin{tabular}{l c c| c}  
  \hline
 \rm{log(g)}  &  $T_{\mathrm{eff}}$ & $[Fe/H]$ & $Reference$\\ \hline
   4.565&   --- & -0.0  &  \citet{kri66}  \\ 
   4.61&   5020 & -0.31  &  \citet{her74}  \\ 
   4.4&   5000  & -0.19  &  \citet{oi74} \\ 
   4.5&   5000 & ---  & \citet{kel78}  \\ 
  4.5&   5100 & ---  &  \citet{tom80}  \\ 
   4.1&  5040 & -0.20  &  \citet{ste81} \\ 
  4.8&  5000 & -0.08  & \citet{bur83}  \\ 
   4.19 &    5040 & -0.23  &  \citet{ste83} \\ 
  4.80&    4990  & -0.20 &  \citet{ab88}  \\ 
 4.61&   5156 & 0.05  &  \citet{bel89}  \\ 
   4.75&   5180 & -0.09 &   \citet{dra93} \\ 
   4.75 &   5000  & 0.06  &  \citet{mal98} \\ 
   4.38 &  5110 & -0.14 &  \citet{tom99} \\ 
   4.57 &  5104 & -0.12 & \citet{zha02} \\
   4.7  &  5135 & -0.07 & \citet{bo03}  \\
   4.62 &  5052 & -0.06 & \citet{al04} \\
   \end{tabular}
   \end{minipage}
   \label{tab:g}
   \end{table}

\subsection{The model}
To build the atmospheric models for $\varepsilon$ Eri, as a first step we computed a photospheric structure capable of reproducing the observed continuum spectrum for this star.
Once the photospheric model was obtained, we changed the chromospheric structure to fit the \ca\ K and H$\beta$ lines for both situations of interest, {\it i.e.} the maximum and minimum levels of chromospheric activity. This is the first time this sort of analysis is made. 

\begin{figure}
\centering
\includegraphics[%
  clip,
  scale=0.4]{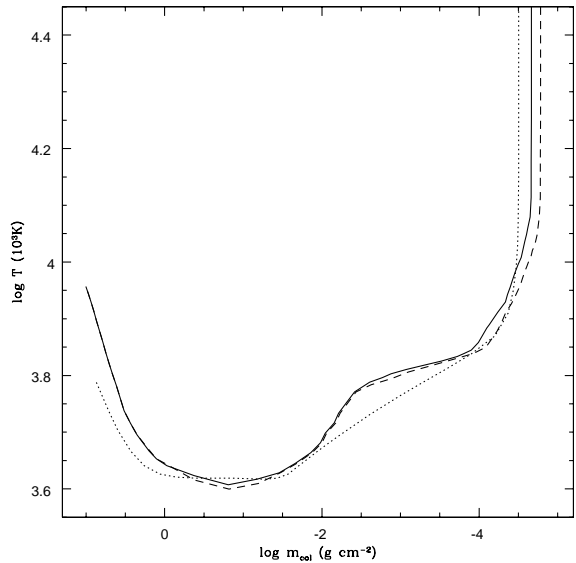}
\caption{\rm Models for $\varepsilon$ Eri in its minimum (dashed line) and its maximum situation 
(full line). For comparison, we show the model B from \citet{sim05} (dotted line)}
\label{erim}
\end{figure}

\begin{figure}
\centering
\includegraphics[%
  clip,
  scale=0.4]{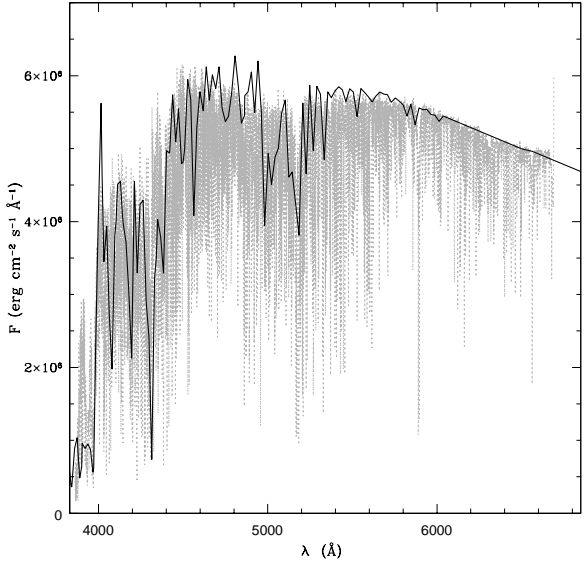}
\caption{\rm Comparison between the observed (grey) and computed continuum (black) for 
$\varepsilon$ Eri.}
\label{cont}
\end{figure}

\begin{figure}
\centering
\includegraphics[%
  clip,
  scale=0.4]{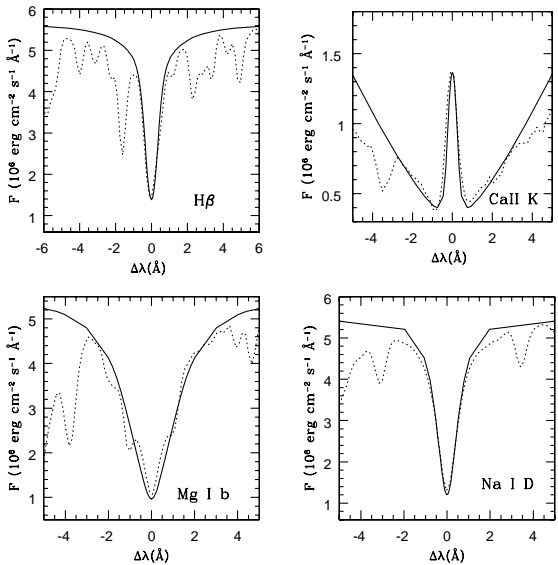}
\caption{\rm Comparison of observed (dashed line) and computed profiles (full line) for 
$\varepsilon$ Eri in its minimum.}
\label{erimi}
\end{figure}

\begin{figure}
\centering
\includegraphics[%
  clip,
  scale=0.4]{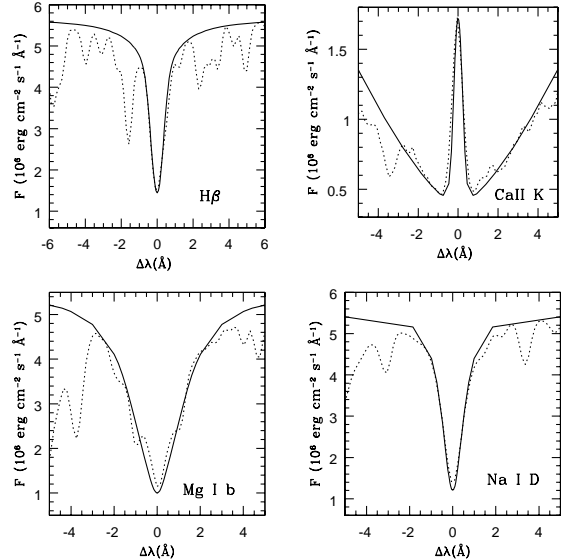}
\caption{\rm Comparison of observed (dashed line) and computed profiles (full line) for 
$\varepsilon$ Eri in its maximum.}
\label{erima}
\end{figure}

Figure~\ref{erim} shows the resulting models, which are presented in column mass for comparison with the best one-component model from SJ05 (their model B). In Figure~\ref{cont} we compare the computed and observed continuum spectrum of $\varepsilon$ Eri, and in Figure~\ref{erimi} and Figure~\ref{erima} we show the comparison of the observed and computed profiles for both levels of activity. It is important to note the good agreement of the fit, even better than the one by \citet{that91} for all the diagnostic lines and continuum.

\textbf{On top of the chromosphere, we added a transition region with a similar 
structure to the solar one. Since we have no observations of lines 
formed in this region, we could no constrain it further. However, the 
position at which the transition region begins was adjusted to fit the observed emission of the \ca\ k line. }

There are several differences between our model and the one by SJ05. In our model, the photosphere is hotter, the temperature minimum region is narrower and the chromospheric rise has a larger slope. Also, their transition region is placed deeper in the atmosphere, {\it i.e.}, at higher values of the column mass.

The differences between these models may be caused by several factors. As we have already noted, SJ05 used a higher value of surface gravity which could explain the differences in all the thermal structure. The differences in the photospheric structure could arise from the fact that we used the complete spectra to fit the continuum emission, and SJ05 used the photospheric model by \citet{that91}, built to fit only the \ca\ K line wings, which are formed in the higher photosphere.

Another important factor to consider is the moment of the activity cycle in which the observations used to build the model were taken. In our case, all the lines used as diagnostics correspond to the same activity level since they were all observed simultaneously. But in the model by SJ05, the structure of the higher chromosphere and transition region was assembled with the model by \citet{that91} for the lower chromosphere 
and photosphere, whitout taking into account that these thermal structures were obtained using line profiles that correspond to different parts of the activity cycle. For these reasons, the comparison between these models is only qualitative.

Regarding the differences in the atmospheric structure between the maximum and the minimum level of activity, the changes occur all along the atmosphere (Figure~\ref{erim}), from the temperature minimum to the transition region. The position of the minimum is the same in both situations, although the temperature increases from 3980 K to 4050 K.

\textbf{Finally, to check whether our results are affected by the adopted value 
of the metallicity, we computed the emitted profiles for our models with 
the metallicity given by \citet{zha02}, which was used by \citet{nes08}, and we found no significant differences. This result is consistent with the one obtained by \citet{ay76}.}

\subsection{The other stars}

To build the models for the other stars in the sample, we used solar metallicity and the same surface gravity that was used for $\varepsilon$ Eri. The stellar surface flux was computed with Equation \ref{flux}, using for each star the \teff\ values shown in Table \ref{tab:ch}.  
Since we want to study the changes in thermal structure induced by activity, we made the approximation that all the stars have the same photosphere than $\varepsilon$ Eri.

\begin{figure}
\centering
\includegraphics[%
  clip,
  scale=0.4]{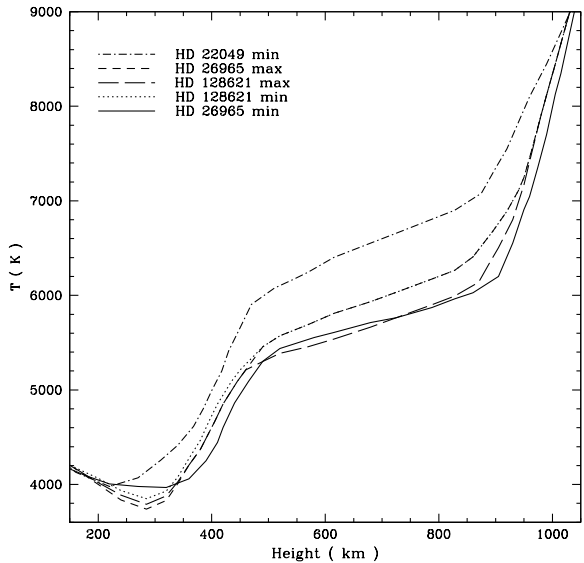}
\caption{\rm Models for the less active group. All the models have the same structure below 150 km.}
\label{menos}
\end{figure}

The models for the less active stars (HD 128621 and HD 26965 in its maximum and minimum activity level, and $\varepsilon$ Eri in its minimum) are shown in Figure~\ref{menos}. It can be seen that all these models have the temperature minimum between 60 and 100 km higher, and from 20 to 240 K cooler that $\varepsilon$ Eri in its minimum.

The temperature in the chromosphere, from the temperature-minimum region up to 1100 km, increases with activity, although the largest differences are in the chromospheric plateau. 
These changes with activity are different to those obtained for G stars (Paper I) with similar activity levels, because in that case only the temperature minimum region changed, and the rest of the atmospheric structure remained the same. 

An important fact which can be seen in Figure \ref{menos}, is that the differences in the atmospheric structure for a star in its maximum and minimum activity levels are comparable to the changes seen beetwen two different stars. This fact stresses how important it is, when building an atmospheric model, the moment at which the observations to be ajusted are made, and, in particular, how important it is to use simultaneous observations of the diagnostic lines.

\begin{figure}
\centering
\includegraphics[%
  clip,
  scale=0.4]{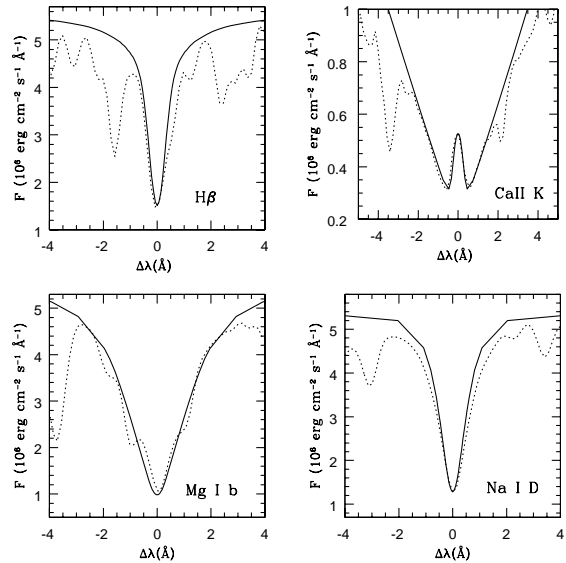}
\caption{\rm Comparison of observed (dashed line) and computed profiles (full line) for 
$\alpha$ Cen B (HD 128621) in its maximum.}
\label{128ma}
\end{figure}

\begin{figure}
\centering
\includegraphics[%
  clip,
  scale=0.41]{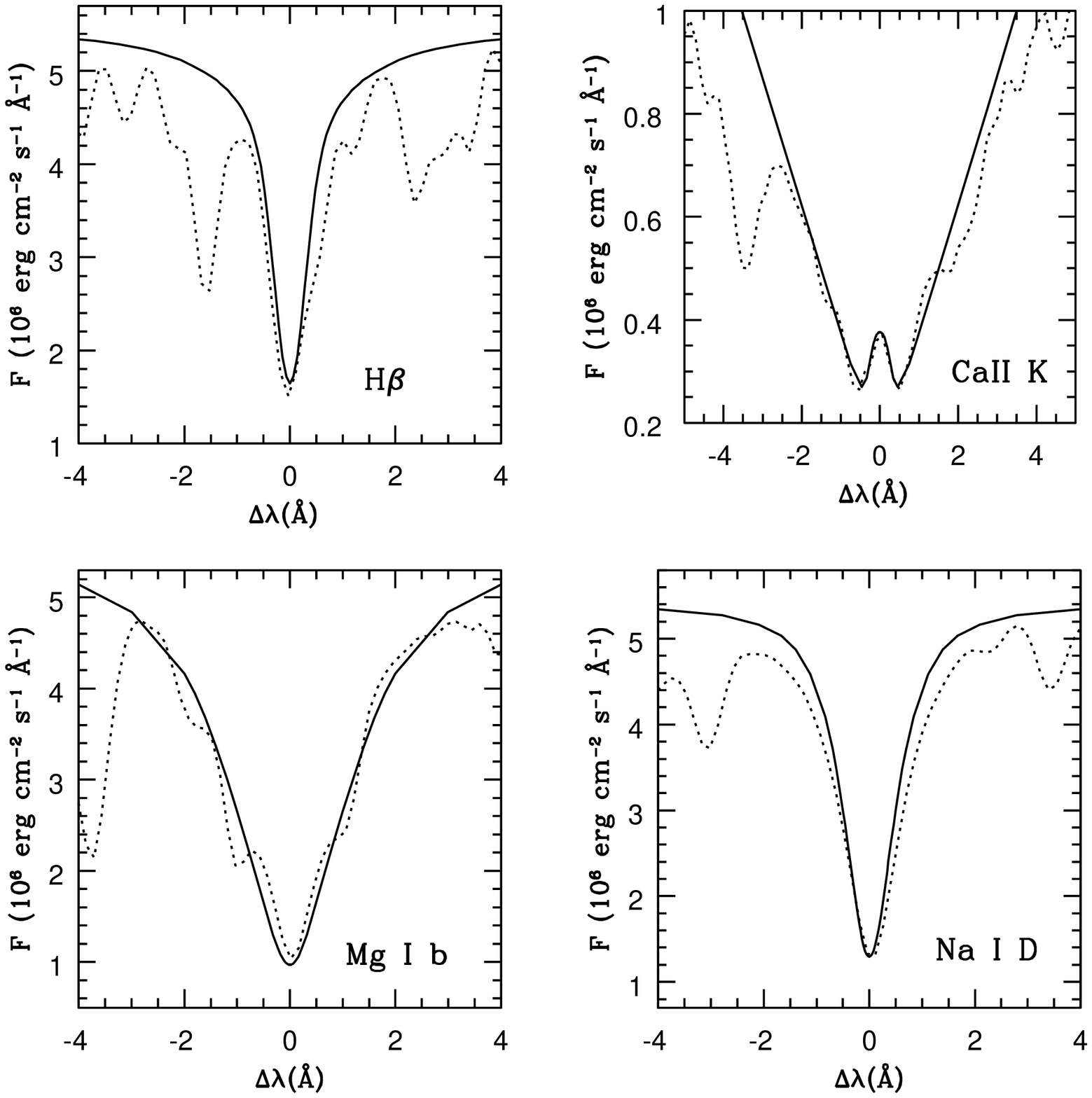}
\caption{\rm Comparison of observed (dashed line) and computed profiles (full line) for 
 $\alpha$ Cen B (HD 128621) in its minimum.}
\label{128mi}
\end{figure}

\begin{figure}
\centering
\includegraphics[%
  clip,
  scale=0.4]{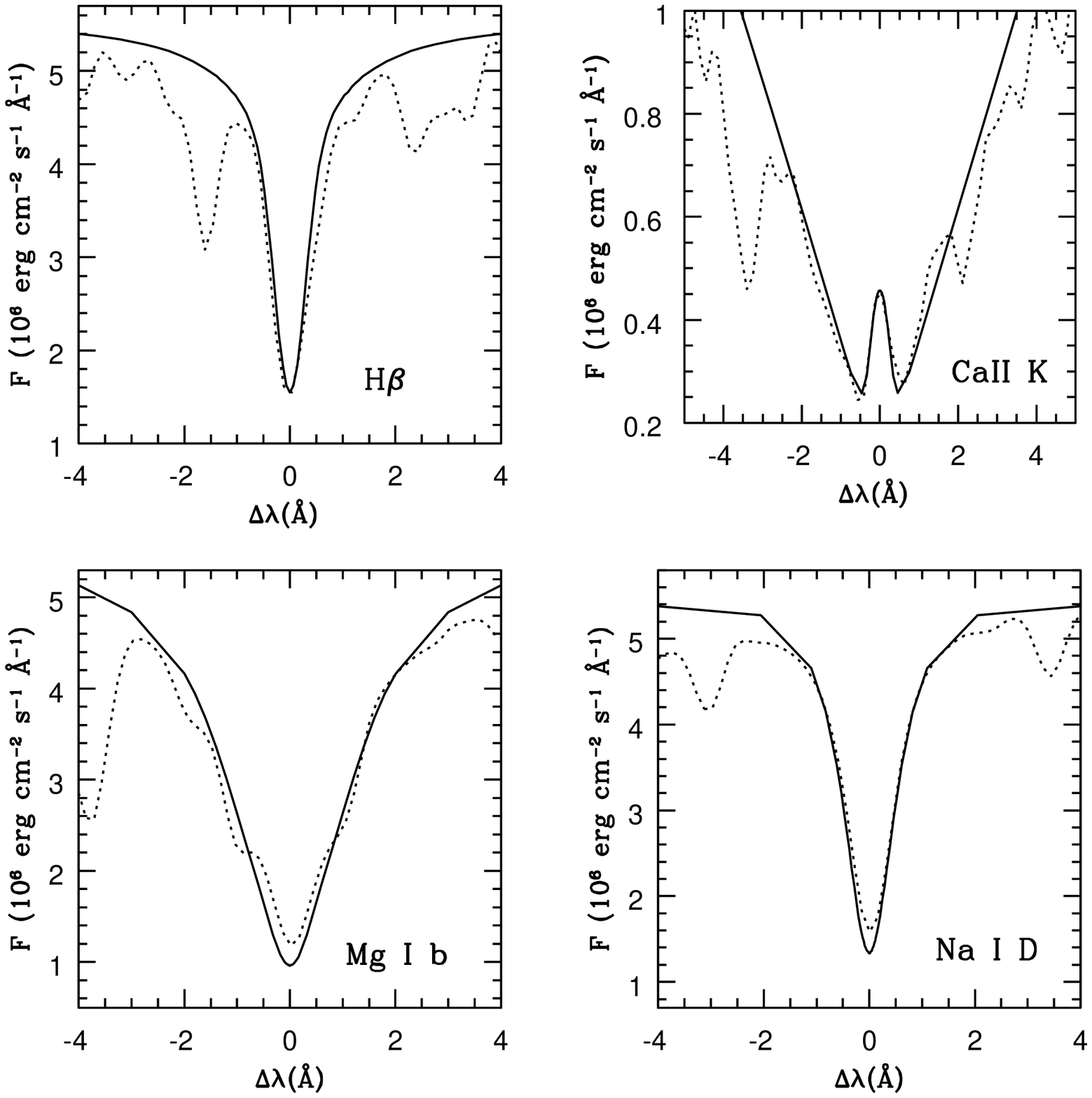}
\caption{\rm Comparison of observed (dashed line) and computed profiles (full line) for 
HD 26965 in its maximum.}
\label{269ma}
\end{figure}

\begin{figure}
\centering
\includegraphics[%
  clip,
  scale=0.4]{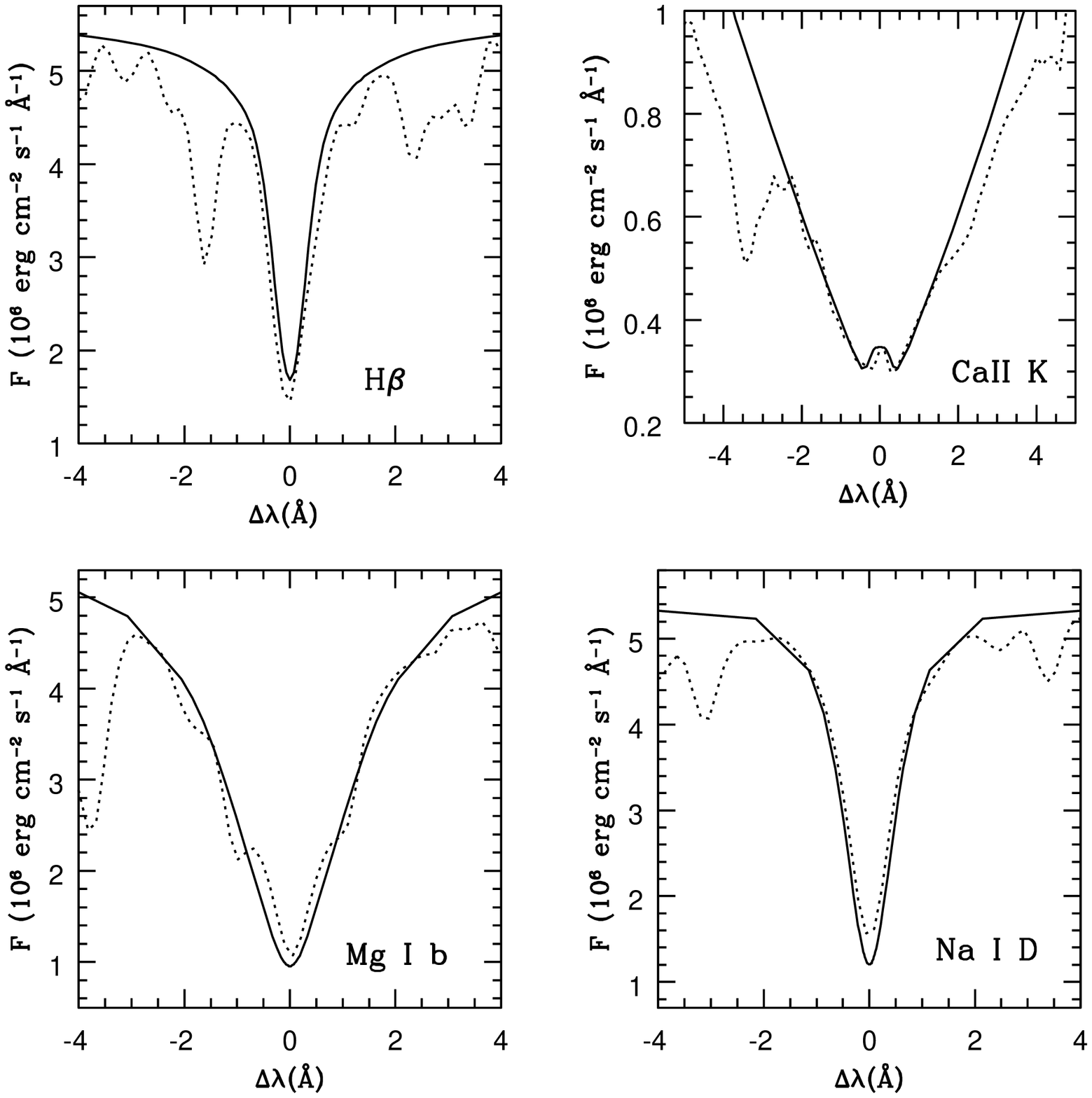}
\caption{\rm Comparison of observed (dashed line) and computed profiles (full line) for 
HD 26965 in its minimum.}
\label{269mi}
\end{figure}

In Figure~\ref{128ma} to \ref{269mi} we show the observed and computed profiles for $\alpha$ Cen B (HD 128621) and HD 26965 in its maximum and minimum states. It is important to note the change in scale to compare with Figures~\ref{erimi} and \ref{erima}, since these two stars are less active than $\varepsilon$ Eridani.

\begin{figure}
\centering
\includegraphics[%
  clip,
  scale=0.4]{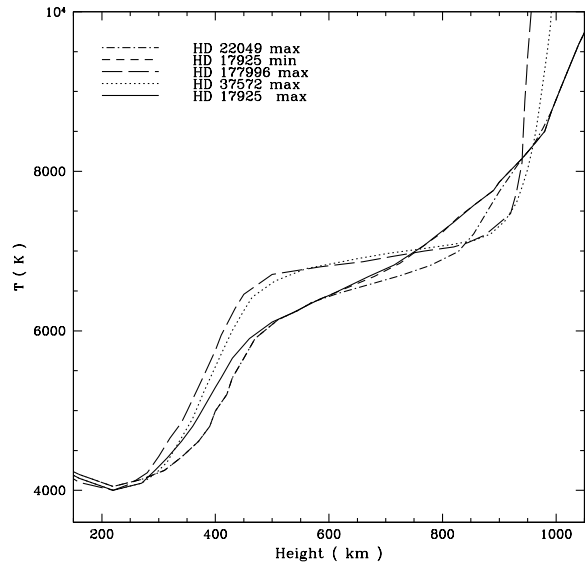}
\caption{\rm Models for the more active group. All the models have the same structure below 150 km.}
\label{mas}
\end{figure}

The models for the most active stars (HD 17925 in both activity levels, and HD 22049, HD 37572 and HD 177996 in their maximum) are shown in Figure~\ref{mas}.
Again the differences in the atmospheric structure for a star in its maximum and minimum activity levels are similar to the changes seen beetwen two different stars.

In Figure~\ref{mas} it can be seen that for the stars in this group the temperature minimum is hotter than for the stars in Figure~\ref{menos}, and this temperature is almost constant as the activity level increases, varying only 50 K. The position of this region is also the same for all these stars.
The atmospheric structure changes with activity everywhere in the chromosphere, mainly in the plateau and the rise to the transition region.

\begin{figure}
\centering
\includegraphics[%
  clip,
  scale=0.4]{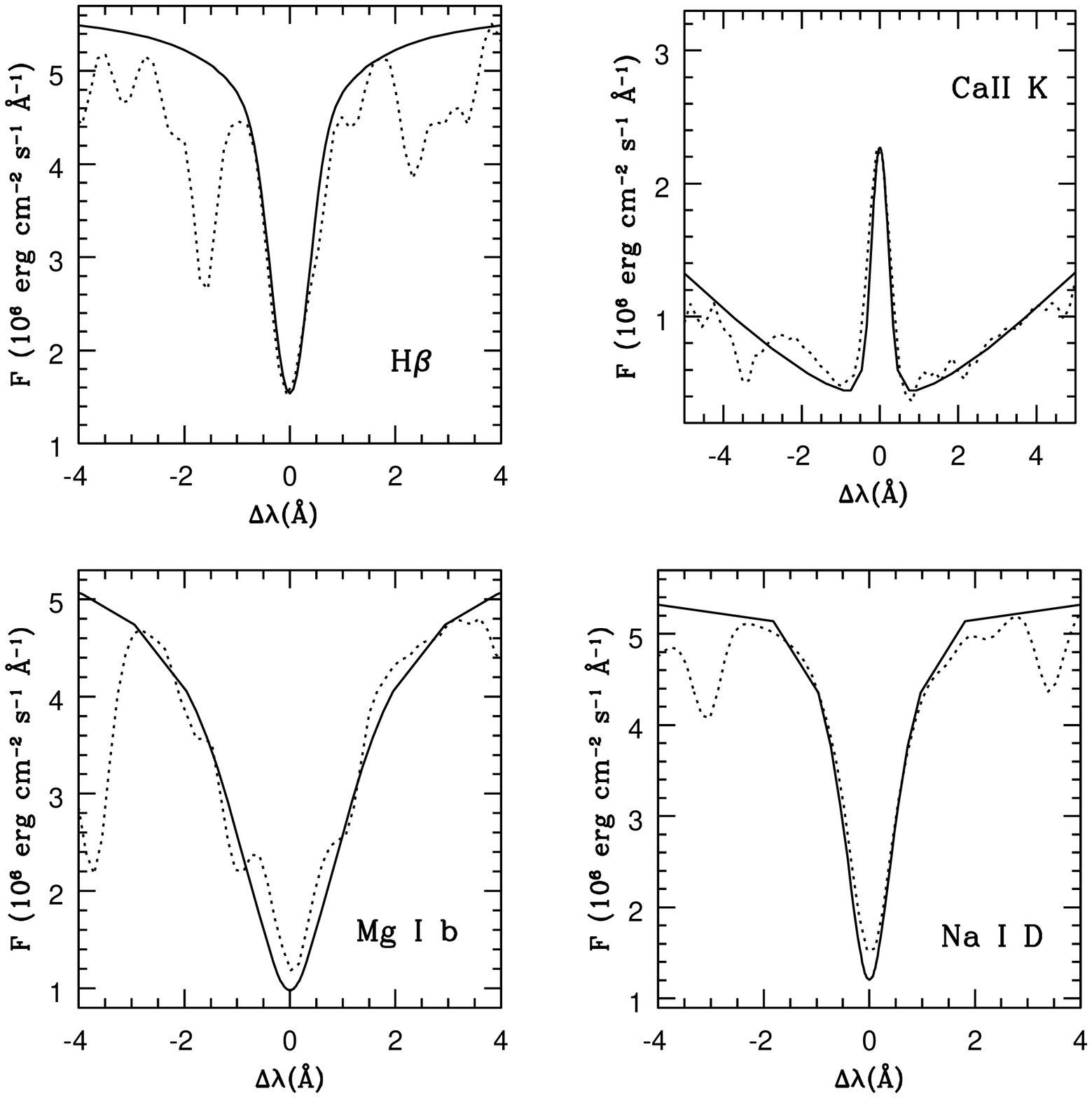}
\caption{\rm Comparison of observed (dashed line) and computed profiles (full line) for 
HD 17925 in its maximum.}
\label{179ma}
\end{figure}

\begin{figure}
\centering
\includegraphics[%
  clip,
  scale=0.4]{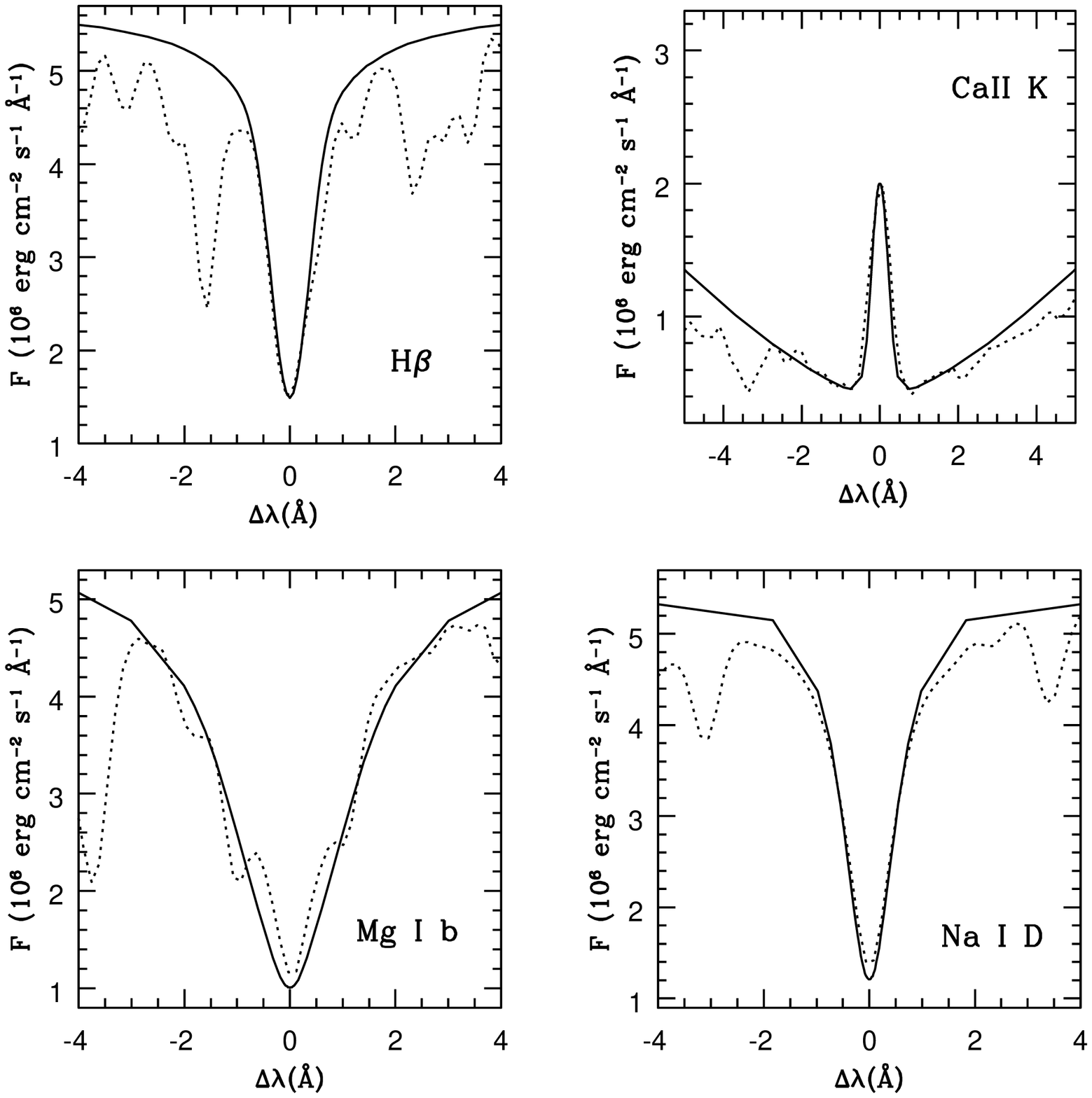}
\caption{\rm Comparison of observed (dashed line) and computed profiles (full line) for 
HD 17925 in its minimum.}
\label{179mi}
\end{figure}

\begin{figure}
\centering
\includegraphics[%
  clip,
  scale=0.4]{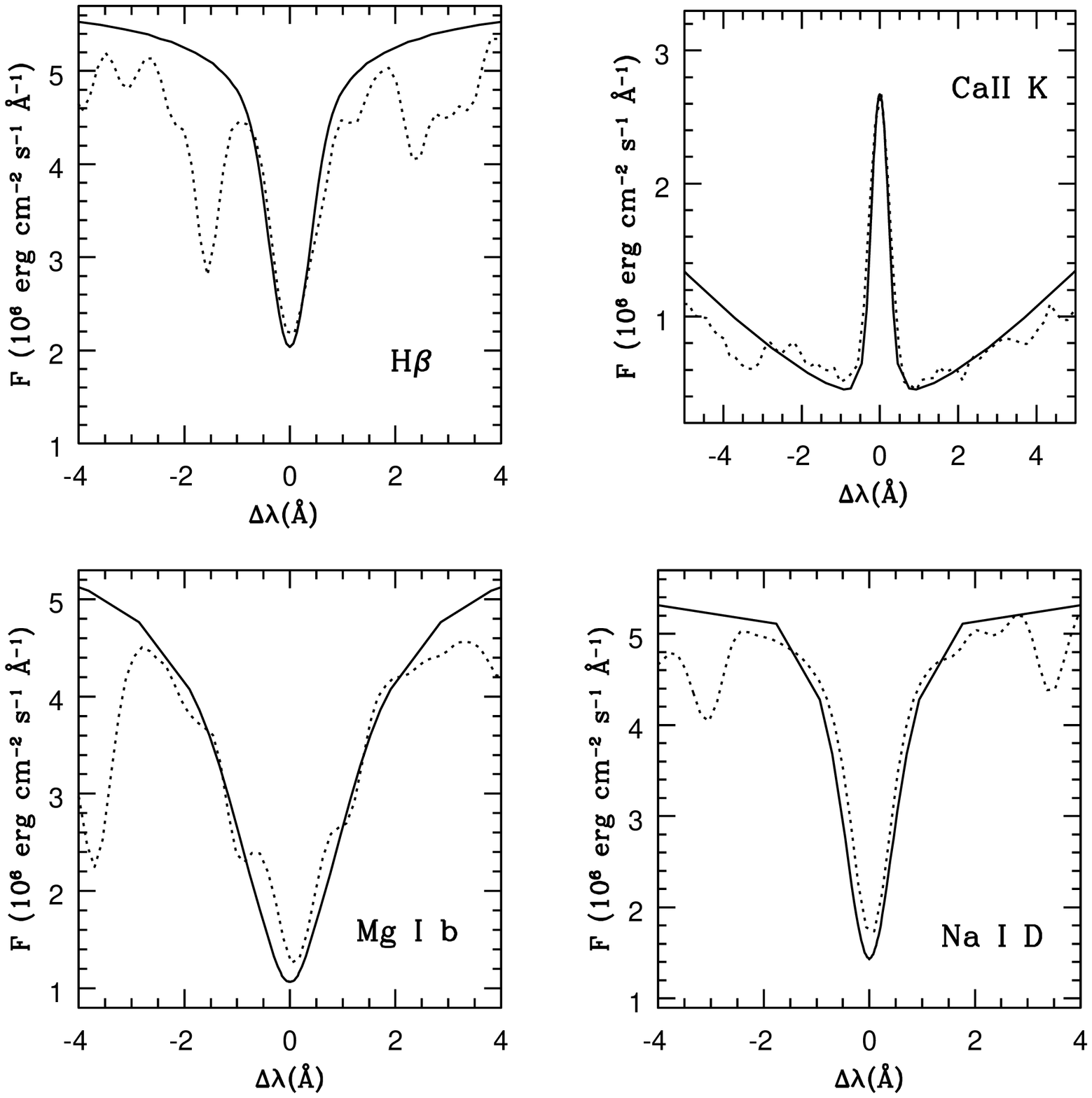}
\caption{\rm Comparison of observed (dashed line) and computed profiles (full line) for 
HD 37572 in its maximum.}
\label{375ma}
\end{figure}

\begin{figure}
\centering
\includegraphics[%
  clip,
  scale=0.4]{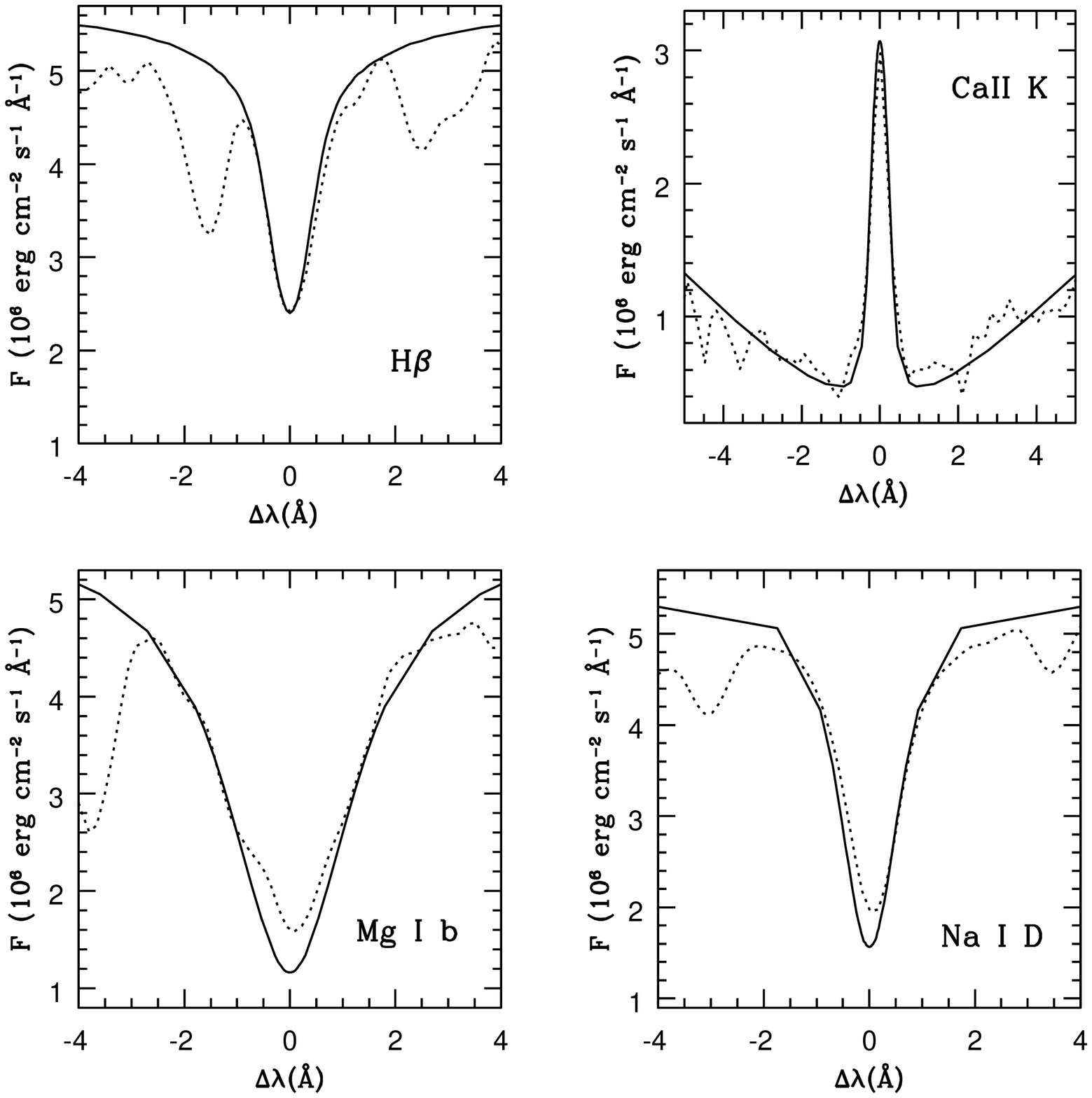}
\caption{\rm Comparison of observed (dashed line) and computed profiles (full line) for 
HD 177996.}
\label{177ma}
\end{figure}

The observed and synthetic profiles for the most active stars are compared in Figures~\ref{179ma} to \ref{177ma}. Again, it is important to note the change in the scale for comparison with the less active stars and the good fit in all cases.

%______________________________________________________________
\section{Non-radiative heating in K stars}

As was mentioned in \citet{kel79}, the ratio of the temperature in the minimum and the effective temperature ($T_\mathrm{{min}}$/$T_\mathrm{{eff}}$) gives an indication of the importance of nonradiative heating in the upper photosphere of stars. In that paper, they compare this ratio with $T_\mathrm{{eff}}$ to study the trend due to spectral type.

\begin{figure} 
\centering
\includegraphics[%
  clip,
  scale=0.35]{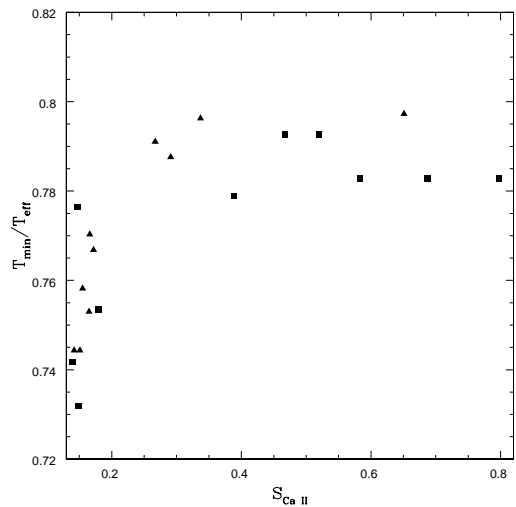}
\caption{\rm  $T_\mathrm{{min}}$/$T_\mathrm{{eff}}$ vs. \Sc\ \textbf{computed from the models} for K stars (this paper, squares) and for G stars (Paper I, triangles).}
\label{tmin}
\end{figure}

In Figure~\ref{tmin} we plot this ratio versus \Sc, which is an indicator of the level of magnetic activity in the chromosphere for all stars independently of spectral type. The values of \Sc\ were obtained by integration of the synthetic profiles, and in the figure we include the values obtained from the models for K stars built in this paper and those for G stars constructed in Paper I.

In the figure it is possible to observe that there is a saturation in $T_\mathrm{{min}}$. In fact, its value increases with activity up to $T_\mathrm{{min}}$/$T_\mathrm{{eff}}$ $\sim 0.79$, and after that it remains almost constant even if activity increases further. On the other hand, the computed value of $T_\mathrm{{min}}$/$T_\mathrm{{eff}}$ for G stars is larger than for K stars with similar activity levels.

\begin{figure*}
\begin{center}
\begin{tabular}{cc}
\hspace{-5.mm}
\psfig{figure=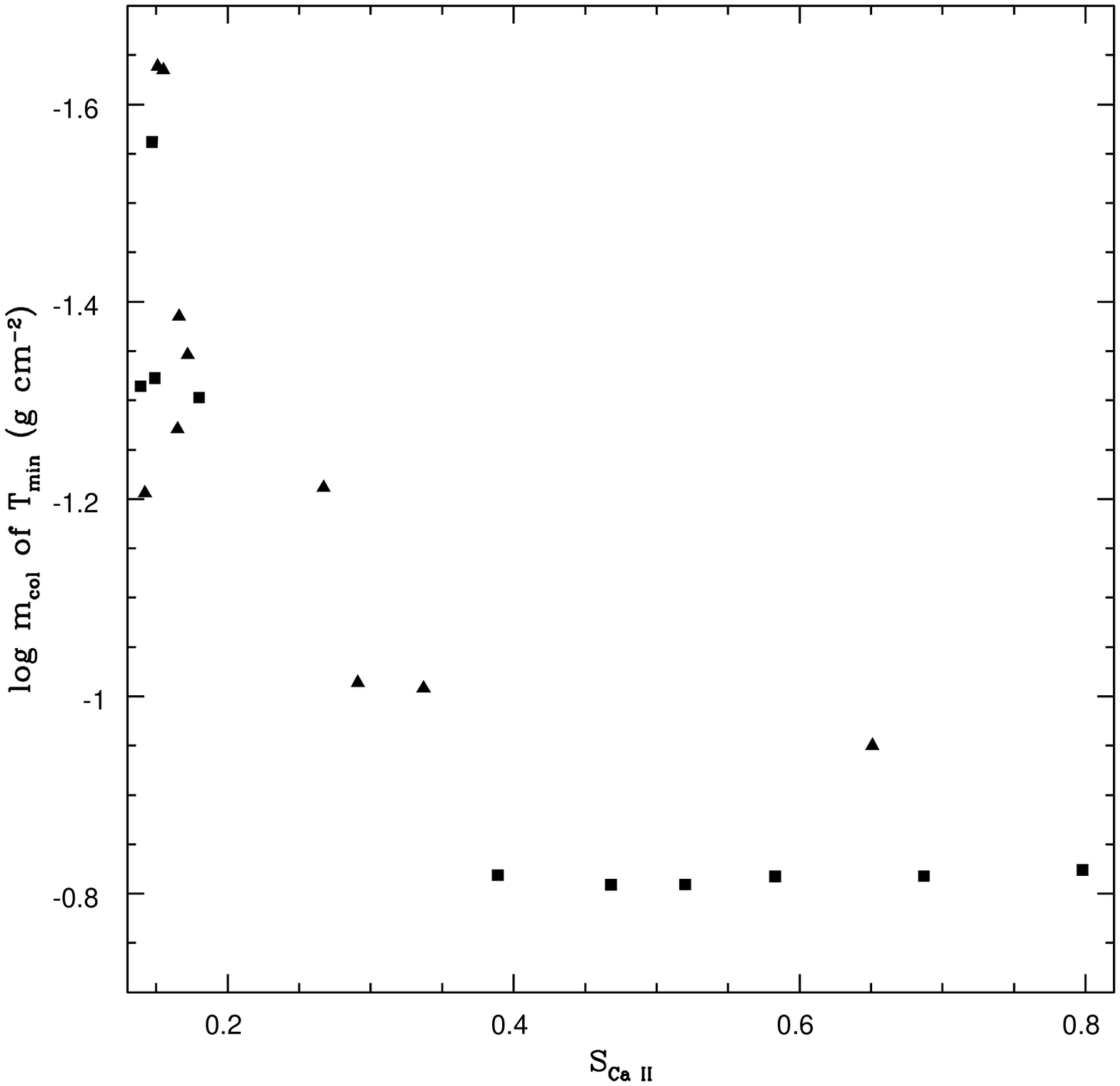,width=80mm,angle=0}
\hspace{0.mm}
\psfig{figure=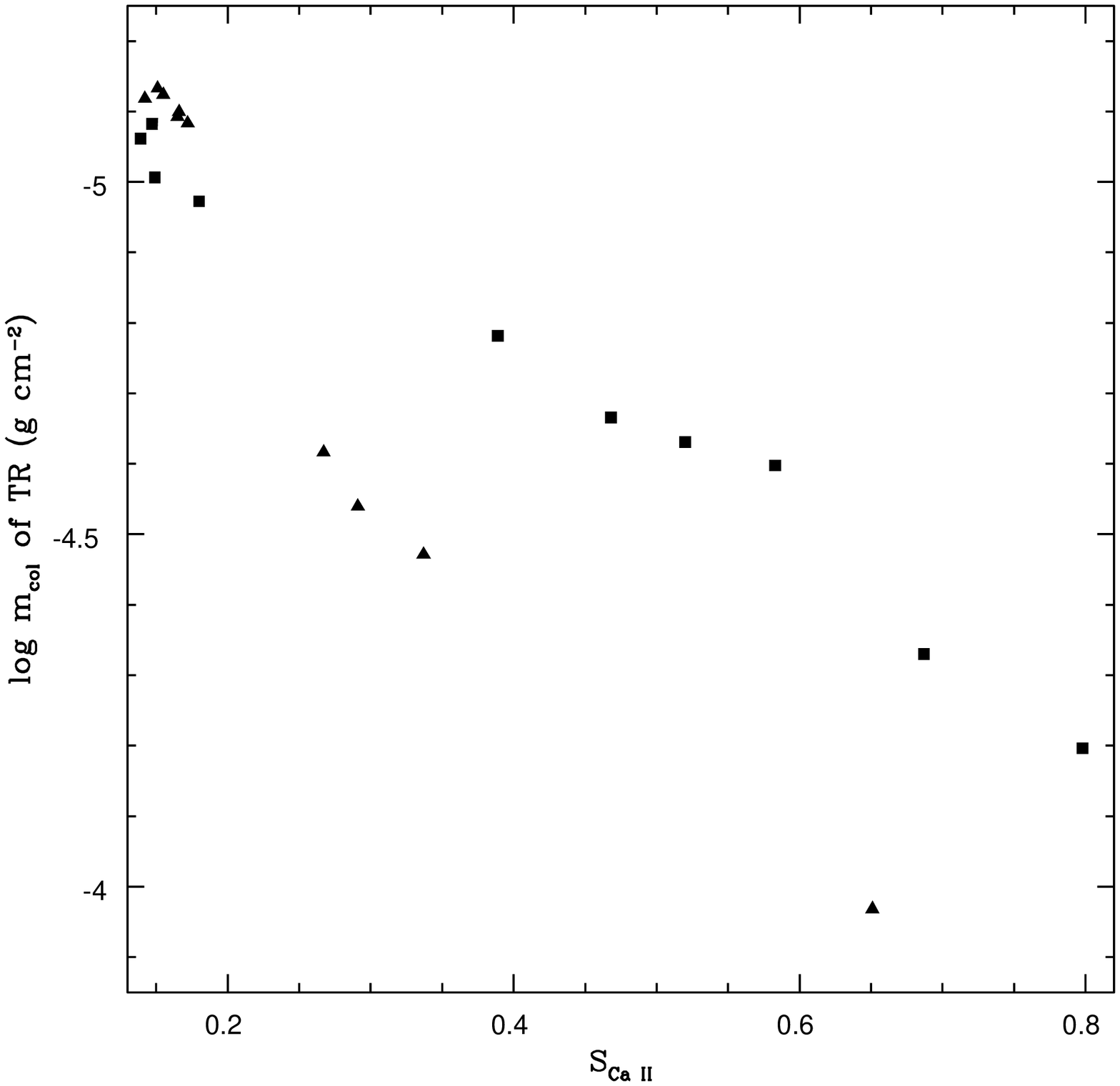,width=80mm,angle=0}
\end{tabular}
\caption{ \rm Position (in column mass) of the temperature minimum region (left) and the transition region (right), as a function of \textbf{computed} \Sc\  for G (Paper I, triangles) and K stars (squares).}
\label{lmin}
\end{center}
\end{figure*}

In Figure~\ref{lmin} (left) we show the position of the temperature minimum region in column mass as a function of \Sc\ for G (Paper I, triangles) and K stars (squares). For K stars the temperature minimum occurs deeper than for G stars, and there is a tendency for this region to move inward as activity grows. In other words, the temperature inversion occurs deeper for more active stars, indicating that the energy deposition starts deeper in the atmosphere as the activity level increases, for both spectral types. 
In Figure~\ref{lmin} we also show the position of the transition region (TR), specifically the height at which the temperature reaches 36000 K. It can be seen that for G stars the chromosphere is more extended than for K stars, and that in both cases the TR moves inward as activity increases.

To study the energetic requirements to maintain the atmospheric structure, we calculated the total 
net radiative loss for each model in the same way as in Paper I. 
At a given depth, the radiative cooling rate $\Phi$ (ergs\, cm$^{-3}$ sec$^{-1}$)
in a given spectral feature (line or continuum) can be computed as (Vernazza
et al. 1981)

\begin{equation}
\Phi= 4\pi \int\kappa_{\nu} \left(S_{\nu}-J_{\nu}\right)  d\nu\,, 
\end{equation}
where $S_{\nu}$ is the source function and $J_{\nu}$ is the mean intensity at frequency 
$\nu$. A positive value of $\Phi$ implies a net loss of energy (cooling), and a negative value
represents a net energy absorption.

\begin{figure}
\centering
\includegraphics[%
  clip,
  scale=0.35]{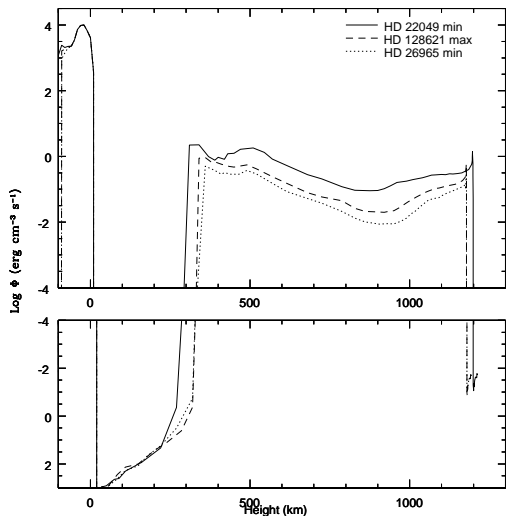}
\caption{\rm Logarithm of the total cooling rate for the less active stars. A
positive value of log $\Phi$ implies a net loss of energy (cooling), and a
negative value represents a net energy absorption.}
\label{inact}
\end{figure}

\begin{figure}
\centering
\includegraphics[%
  clip,
  scale=0.35]{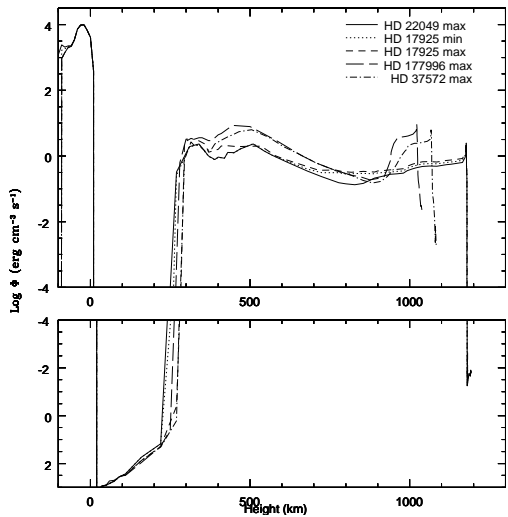}
\caption{\rm Logarithm of the total cooling rate for the more active stars. A
 positive value of log $\Phi$ implies a net loss of energy (cooling), and a
 negative value represents a net energy absorption.}
\label{act}
\end{figure}

Here, we considered line and continua of H, H-, H-ff, Mg I and II, Fe
I, Si I, Ca II, Na I and CO.  
The total rates for each star are shown in Figure~\ref{inact} for the
less active models, and in Figure~\ref{act} for the more active
ones. As it is expected, the amount of non-radiative energy supplied
to the chromosphere increases everywhere with magnetic activity.  

In both figures, it is possible to note a region where the net cooling
rate is negative. This fact was already known for the Sun (Vernazza et
al. 1981), and was later found in Paper I for other G stars,
for which negative cooling rates in the temperature minimum
region were also obtained.  
Within the plane-parallel, homogeneous approximation we are
investigating, this implies either mechanical energy extraction or,
more likely, that the calculations have neglected important sources of
radiative cooling (see \citeauthor{ma}1993). 

The main contributions in this zone are H-, Si I, Fe I and CO, the
same than for G stars. It is important to note that since the
temperature for K stars is lower in this region, there could be an
important contribution of several molecules which we do not consider
in our calculations, like, for example, CH, that could act as cooling
agents. Considering these contributions could bring our computations
closer to energy balance. 

For the less active models the cooling rate becomes positive at around
300 km, implying that there is mechanical energy deposition above this
height. For the most active models, this energy deposition starts
deeper in the atmosphere, {\it i.e.} the chromosphere starts deeper.

Also in the chromosphere, the most important contributors to the cooling rate are the same than for G stars, but the proportions are 
different: for $\varepsilon$ Eri in its minimum, for example, Mg II
and Ca II contribute with $\sim 9 \%$ each, while for the Sun these
contributions are of $\sim 20 \%$. The contribution by Fe I, on the
other hand, is of $\sim 15 \%$ in $\varepsilon$ Eri, but only $\sim 10
\%$ in the Sun. In both cases, almost half of the total cooling rate
corresponds to line blanketing.  

\begin{figure}
\centering
\includegraphics[%
  clip,
  scale=0.35]{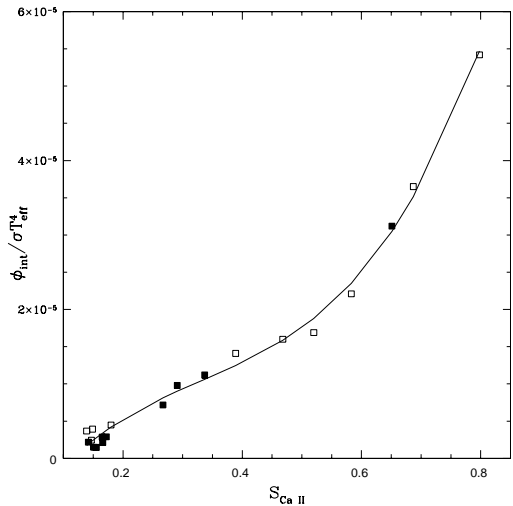}
\caption{\rm Normalized $\phi_\mathrm{{int}}$ versus \textbf{computed}
  \Sc\ index.  \textbf{Empty} squares represent the K star models from
  this work and \textbf{full squares} indicate the G star models
  from Paper I.} 
\label{svsco2}
\end{figure}

Finally, to quantify the total amount of mechanical energy deposited
in the chromosphere, we integrated the net radiative cooling rate 
from the depth in the chromosphere where the cooling rate becomes positive
to the region where the temperature reaches $10^{4}$ K. To compare the
results for both spectral types, we normalized the integrated rate,
$\phi_\mathrm{{int}}$, by the surface luminosity ($\sigma
T_{\mathrm{eff}}^{4}$) \textbf{The resulting quantity, therefore,
gives an idea of the fraction of the total energy emitted by the star
that goes into heating the chromosphere.}    
The results are shown in Figure~\ref{svsco2}, where it can be seen
that there is a unique trend for all stars, independently of
spectral type. This fact seems to imply that the physical processes
that supply the energy to sustain the atmospheric structure are
independent of spectral type. 

%Figure~\ref{svsco} shows the resulting quantity,
%$\phi_\mathrm{{int}}$, versus \Sc\ computed
%from the models, for the K stars (triangles) and for the G stars from
%Paper I (squares). In this figure, the amount of non-radiative energy
%needed to heat the chromosphere, for the same range of stellar
%activity, is larger in G stars than in K stars.  

%\begin{figure}[htb!]
%\centering
%\includegraphics[%
%  clip,  scale=0.35]{svsco.eps}
%\caption{\rm Integrated cooling rate versus \Sc ~index. The square are the K star models from this 
%work and the triangle are G star models from Paper I.}
%\label{svsco}
%\end{figure} 

Given the good corelation between \Sc\ and the normalized
$\phi_\mathrm{{int}}$, we fit the data with a polynomial function,
given by 
%\begin{equation}
\begin{eqnarray}
\frac{\phi_\mathrm{{int}}}{\sigma T_{\mathrm{eff}}^{4}}= -1.14 ~10^{-5} + 1.28 ~10^{-4}~ \rm{S}_{Ca II}
\nonumber \\ 
+ 2.80  ~10^{-4} ~\rm{S}_{Ca II}^{2} - 2.80 ~10^{-4} ~\rm{S}_{Ca  II}^{3}\ . 
\label{emp}
\end{eqnarray}
%\end{equation}
In Figure~\ref{svsco2} it can be seen that the fit is very good and,
therefore, the energetic requirements of a given star can be estimated from its
chromospheric activity level as mesured by \Sc. 

%______________________________________________________________

\section{Discussion}

One of the main goals of  chromospheric modelling is to accurately
estimate the radiative losses in the chromosphere in detail, using
only the information that can be obtained from the observations,
without any assumption about the physical processes involved. In this
way, these losses can be equaled to the energy requirements that any
proposed mechanism of chromospheric heating should match.

%\cite{cun99}~(1999) computed a theoretical two-component model for K
%dwarfs of different activity levels and compared the results with
%observation. As part of their stellar sample, there are  two stars of
%ours, $\varepsilon$ Eri and HD 17925. The two components are a
%nonmagnetic one heated by acoustic waves and a magnetic component
%heated by longitudinal tube waves. To compare with observations they
%calculate the radiative losses from the chromosphere. They solved the
%radiative transfer equation with partial %redistribution together
%with the statistical equilibrium for NLTE, but considered only the
%contribution of Mg II h and k line, as the responsible for the
%radiative losses. Then, when the full balance is established between
%the dissipated energy from the two theoretical components and the
%local radiative losses from the radiative transfer calculation, new
%values of atmospheric temperature and density are calculated giving
%the time-dependent atmospheric model.

\textbf{For example, we saw in the previous section that the contribution
of the different features to the total cooling rate is not the same for G and
K stars. In particular, the Ca II, Mg II and Fe I relative
contributions are not the same for both spectral types. Therefore, it
might not be correct to scale the relative contributions computed for
the Sun to K stars, as it has been done sometimes 
% However, in previous works on chromospheric heating have been
% considered that the total losses are the same than in the solar
% model C of \citet{val} for stars of different spectral type 
(see \citeauthor{cun99} \citeyear{cun99}, and \citeauthor{ram} \citeyear{ram}).} 

% in K stars is half
%that of the Sun, and therefore Cuntz et al.~(1999) underestimated the
%total rate by a factor of 2. 

%For example, Cuntz et al. (1999) computed theoretical two-component
%models for K dwarfs of different activity levels.
%To establish the energetic balance, they computed the chromospheric
%radiative losses in the Mg II k line in detail, and obtained the total
%losses assuming that each spectral feature contributed in the same
%proportion than in the solar model C of \citet{val} (model VALC
%1981; see also \citeauthor{ram}2005). However, 

Cuntz et al. (1999) computed theoretical two-component
models for K dwarfs of different activity levels.
They proposed that the energy is deposited in the
chromosphere by acoustic and magnetic shocks, and found that these
shocks are stronger and are produced deeper in the chromosphere as the
activity of the star increases. 
This result is in agreement with our calculations, which shows that
the energy deposition is larger and deposited deeper with increasing
activity.

On the other hand, they reproduced the lineal trend between the \ca\ H
and K lines fluxes and the rotational period, although their 
computed fluxes are smaller than the observations, which could 
be due to their sketchy calculation of the radiative cooling rate. 

\section{Summary}

In this paper we present chromospheric models for six K dwarfs, including $\varepsilon$
Eridani, with similar photospheric properties 
but different magnetic activity levels. In most cases we computed
models for two moments of the activity cycle for the same star. 

These models were based on, and reproduced very well, the Ca II H and
K and the H$\beta$ line profiles for all the stars in our sample. The
reliability of the stellar atmospheric models was checked with other
features, the Na I D and Mg I b lines. Also for these lines we found
very good agreement between computed and observed profiles.  

We found that the changes in atmospheric structure in K dwarfs with
activity are produced all along the chromosphere, from the region of
the temperature minimum to the transition region and mainly in the
chromospheric plateau, independently of the activity level of the star.  
This was not the case for the G dwarfs modelled in Paper I, since for the less active G
stars the changes with activity occur only in the region of the
temperature minimum.   
 
The ratio of the minimum and effective temperatures
($T_\mathrm{{min}}$/$T_\mathrm{{eff}}$) can give an idea of the
importance of non-radiative heating in the upper photosphere of stars.  
Both for K and G stars, this value increases with activity up to 
$T_\mathrm{{min}}$/$T_\mathrm{{eff}}$ $\sim 0.79$, where it
saturates, and it remains constant even if the activity level
increases further. On the other hand, the computed value of 
$T_\mathrm{{min}}$/$T_\mathrm{{eff}}$ for G stars is larger than for K
stars with similar activity.

For both spectral types, the position of the temperature minimum moves
inward as activity increases, implying that the chromosphere starts
deeper in more active stars. This, in turn, implies that as the
activity level increases, the energy deposition occurs deeper in the
atmosphere. 

On the other hand, the transition region is placed at higher column
masses for G stars than for K stars, and in both cases it moves
inward as activity increases.  

Regarding the energetic requirements, the integrated chromospheric
radiative losses, normalized to the surface luminosity, show a unique
trend  for G and K dwarfs when plotted against $\rm{S}_{Ca II}$,
the main proxy of stellar activity.  This might indicate that
the same physical processes are heating the stellar chromospheres in
both cases. We calculated an empirical relationship between the \Sc\
index and the energy deposited in the chromosphere, which can be used to
estimate the energetic requirements of a given star knowing its
chromospheric activity level. 

There are significant differences in the contributions of Mg II, Ca II and Fe I to the
total net cooling rate in the chromosphere between G and K stars,
which implies that values obtained for a given star should not be
extrapolated to another one of a different spectral type. 
In both cases about half of the total rate is due to line blanketing.  

%______________________________________________________________ 


%% Included in this acknowledgments section are examples of the
%% AASTeX hypertext markup commands. Use \url without the optional [HREF]
%% argument when you want to print the url directly in the text. Otherwise,
%% use either \url or \anchor, with the HREF as the first argument and the
%% text to be printed in the second.

\section*{Acknowledgments}
We would like to thank the director of the CASLEO Observatory, and all
the staff of this institution. The CCD and data acquisition system at
CASLEO has been partly financed by R.M. Rich through U.S. NSF grant
AST-90-15827.
We also thank the anonymous referee, whose comments help us to
improve this paper.
This work made extensive use of the SIMBAD database, operated at CDS, Strasbourg, France.

\end{document}